\documentclass[9pt,twocolumn,twoside]{pnas-new}

\templatetype{pnasresearcharticle} 

\usepackage{pdfpages}
\usepackage[final]{changes}
\definechangesauthor[name={ACionca},color=magenta]{AC}
\nolinenumbers

\newcommand\SICanonicalModularity{SI Appendix, Fig. S1}
\newcommand\SICanonicalStability{SI Appendix, Fig. S2}
\newcommand\SICelegansPos{SI Appendix, Fig. S3}
\newcommand\SICelegansPathways{SI Appendix, Fig. S4}
\newcommand\SICelegansMoreBicom{SI Appendix, Fig. S5}
\newcommand\SICanonicalSourceSink{SI Appendix, Fig. S6}
\newcommand\SICanonicalDissort{SI Appendix, Fig. S7}
\newcommand\SICelegansSecEmbed{SI Appendix, Fig. S8}
\newcommand\SICanonicalBenchmark{SI Appendix, Fig. S9}
\newcommand\SICelegansGap{SI Appendix, Fig. S10}
\newcommand\SIDatasetCelegans{SI Dataset S1}

\begin{document}

\title{Community detection for directed networks revisited using bimodularity}

\author[a,b]{Alexandre Cionca}
\author[a,b]{Chun Hei Michael Chan}
\author[a,b,1]{Dimitri Van De Ville}

\affil[a]{Neuro-X Institute, École Polytechnique Fédérale de Lausanne (EPFL), Geneva, Switzerland}
\affil[b]{Department of Radiology and Medical Informatics, University of Geneva, Geneva, Switzerland}

\leadauthor{Cionca}

\significancestatement{
    The art of finding patterns or communities plays a central role in the analysis of structured data such as networks.
    Community detection in graphs has become a field on its own.
    Real-world networks, however, tend to describe asymmetric, directed relationships and community detection methods have not yet reached consensus on how to define and retrieve communities in this setting.
    This work introduces a framework for the interpretation of directed graph partitions and communities, for which we define the \textit{bimodularity} index and provide an optimization method to retrieve the embedding and detection of directed communities.
    The application of our approach to the worm neuronal wiring diagram highlights the importance of directed information that remains hidden from conventional community detection. 
}

\authorcontributions{Author contributions: A.C. and D.V.D.V designed the research; A.C, M.C and D.V.D.V performed the research; A.C. and D.V.D.V analyzed data; A.C, M.C and D.V.D.V wrote the article.}
\authordeclaration{The authors declare no conflict of interest.}
\equalauthors{}
\correspondingauthor{\textsuperscript{1}To whom correspondence should be addressed. E-mail: dimitri.vandeville@epfl.ch}


\begin{abstract}
Community structure is a key feature omnipresent in real-world network data. 
Plethora of methods have been proposed to reveal subsets of densely interconnected nodes using criteria such as the modularity index.
These approaches have been successful for undirected graphs, but directed edge information has not yet been dealt with in a satisfactory way.
Here, we revisit the concept of directed communities as a mapping between sending and receiving communities.
This translates into a new definition that we term bimodularity.
Using convex relaxation, bimodularity can be optimized with the singular value decomposition of the directed modularity matrix.
Subsequently, we propose an edge-based clustering approach to reveal the directed communities including their mappings. 
The feasibility of the new framework is illustrated on a synthetic model and further applied to the neuronal wiring diagram of the \textit{C. elegans}, for which it yields meaningful feedforward loops of the head and body motion systems.
This framework sets the ground for the understanding and detection of community structures in directed networks.
\end{abstract}

\dates{This manuscript was compiled on \today}
\doi{\url{www.pnas.org/cgi/doi/10.1073/pnas.XXXXXXXXXX}}

\maketitle
\thispagestyle{firststyle}
\ifthenelse{\boolean{shortarticle}}{\ifthenelse{\boolean{singlecolumn}}{\abscontentformatted}{\abscontent}}{}

\firstpage[3]{3}

\dropcap{N}etworks are powerful models to represent interactions within almost any type of structured data.
Nodes and edges can symbolize relationships between agents in social networks, metabolic interactions between cells, or information flow between neuronal populations in the human brain.
The study of the emerging network properties provides critical insights into the observed data \cite{varshney_structural_2011, barabasi_network_2011, vertes_simple_2012, fornito_connectomics_2015, bassett_network_2017, li_graph_2022}.
One predominant attribute of many real-world networks is the presence of community structure where subsets of nodes are more densely connected between them than expected by the degree distribution \cite{girvan_community_2002}.
Community detection is about finding the best graph partitioning $\mathcal{C}_k$, $k=1,\ldots K$ \cite{fortunato_community_2016, rohe_co-clustering_2016, pathak_hierarchy_2024}.
Mathematically, the modularity index $Q$ for an undirected graph captures the excess proportion of edges running within the communities \cite{bender_asymptotic_1978, newman_structure_2003}:
\begin{equation}
    Q=\frac{1}{2m} \sum_{k=1}^K \sum_{i,j\in C_{k}}\left[A_{ij}-\mathbb{E}(A_{ij}|\mathcal{H}_0)\right],
\end{equation}
where $m$ is the total number of edges, the elements of adjacency matrix $A_{ij}$ contain the edge weights between nodes $i$ and $j$, and $\mathbb{E}(A_{ij}|\mathcal{H}_0)$ expresses the expected proportion under the null hypothesis.
The configuration model is a common null hypothesis that redistributes the edges weights evenly over the nodes; i.e., $k_i k_j / (2m)$ where $k_i$ is the degree of node $i$.
Modularity can thus be seen as a statistical measure of the unexpectedness in edges arrangement and of how graph partitions are exploiting such modular structures to form densely connected (assortative) or bipartite (dissortative) communities \cite{holland_stochastic_1983, snijders_estimation_1997, peixoto_bayesian_2019}.
Many approaches have been proposed to optimize modularity, such as the Louvain algorithm which operates in the graph domain~\cite{blondel_fast_2008, lambiotte_communities_2009} or the spectral method that eigendecomposes the modularity matrix ${\mathbf B}$ where $B_{ij}=A_{ij}-{k_i k_j}/(2m)$ \cite{newman_modularity_2006, newman_finding_2006}.
Such optimizations can be interpreted as embedding problems for the graph nodes or edge, which in the latter may provide overlapping groups of nodes \cite{evans_line_2009, ahn_link_2010}.

At first sight, the definition of modularity is amendable for directed graphs.
First, the adjacency matrix considered in $Q$ can be asymmetric. Second, the null model can be changed to $k^\text{out}_i k^\text{in}_j/m$ to account for out- and in-degree of the nodes \cite{arenas_size_2007, leicht_community_2008}.
However, this type of directed modularity does not consider edge direction \cite{malliaros_clustering_2013, kim_finding_2010}; i.e., the contribution to modularity of a specific edge between nodes $i$ and $j$ remains constant as long as $A_{ij}+A_{ji}$ and the in- and out-degrees are the same \cite{kim_finding_2010}.
In addition, the spectral method cannot be applied to non-symmetric matrices.

Considerable work has been carried out to go beyond these limitations for community detection in directed networks \cite{leicht_community_2008, kim_finding_2010, hosseini-pozveh_label_2022, dang_community_2023, rohe_co-clustering_2016, wang_spectral_2020, sussman_consistent_2012}.
First, the spectral method has been applied to the symmetrized matrix $\mathbf{B}+\mathbf{B}^T$.
However, because this approach obfuscates edge direction \cite{leicht_community_2008}, the directed modularity has instead been redefined with random walk priors in the \textit{LinkRank} method \cite{kim_finding_2010}.
It is defined as the difference between the fraction of time a random walker will spend within communities and the expected value of this fraction.
While this approach accounts for edge direction and is coherent with the undirected modularity, the matrix definition of the \textit{LinkRank} modularity is, however, asymmetric and the authors therefore present an optimization process that again requires symmetrization similar to the one in \cite{leicht_community_2008}.
\added[id=AC]{Alternative symmetrization methods have also been considered, such as the weighted label propagation algorithm (WLPA) to convert a directed network into an undirected weighted yet signed graph \cite{hosseini-pozveh_label_2022}.
Second, recent approaches proposed a workaround to matrix symmetrization through spectral clustering based on the singular value decomposition (SVD) of directed graph operators \cite{dang_community_2023, rohe_co-clustering_2016, wang_spectral_2020, sussman_consistent_2012}.
The left and right singular vectors indeed capture meaningful node embeddings that are tied to the row and column spaces, respectively, of the asymmetric normalized Laplacian \cite{dang_community_2023}, regularized Laplacian \cite{rohe_co-clustering_2016, wang_spectral_2020}, and adjacency \cite{sussman_consistent_2012} matrices.
While the SVD offers relevant adaptations of spectral methods to directed graphs, none of these frameworks combines the SVD with the notion of a null model that is proper to modularity, nor leverage the natural correspondence between left- and right-singular vectors in their clustering scheme.}
In sum, there is not yet a satisfactory\added[id=AC]{, intuitive, and straightforward} extension of community detection to directed graphs.

Here, we revisit the definition of modularity in a subtle but essential way.
In particular, instead of considering a single partitioning in terms of communities, we propose the concept of \textit{bimodularity} that identifies both sending and corresponding receiving communities; i.e., two partitions that are not necessarily overlapping.
The directed nature of the graph can therefore be reflected in terms of community structure.
We also derive an efficient algorithm that jointly optimizes both partitions and detects directed communities.
The feasibility and interpretation are first demonstrated on synthetic examples.
Finally, insightful results are obtained on the directed neuronal wiring diagram of the nematode \textit{Caenorhabditis elegans} (\textit{C. elegans}) and indicate the potential of this new method for real-world applications.

\section*{Results}

\subsection*{Bimodularity}
Communities in undirected graphs are characterized by more edges than expected by the null model, however, in directed graphs, these edges do not necessarily map to the same set of nodes.
Therefore, we introduce $K$ sending communities as $\mathcal{C}^\text{out}=\big(C_{1}^\text{out}, C_{2}^\text{out}, ..., C_{K}^\text{out}\big)$ and assume a mapping $\mathcal{M}$ that relates each sending community to a corresponding receiving one such that $\mathcal{C}^\text{in}=\big(C_{1}^\text{in}, C_{2}^\text{in}, ..., C_{K}^\text{in}\big)$, where $\mathcal{M}(C^\text{out}_k)=C^\text{in}_k$.
The sending and receiving communities can also be overlapping between them.
We now define bimodularity $Q_\text{bi}$ as the difference between the fraction of edges from sending communities to their respective receiving communities and the expected value of this fraction:
\begin{align}
    Q_\text{bi}&=\frac{1}{m}\sum_{k=1}^K \sum_{\substack{i\in C_{k}^\text{out}\\j\in C_{k}^\text{in}}} \left[A_{ij}-\mathbb{E}(A_{ij}|{\mathcal H}_0)\right],
\end{align}
where edges $(i,j)$ run from a sending to the corresponding receiving community.
Bimodularity reverts to conventional modularity when the graph is undirected and the two partitions coincide (see Methods for details).

\begin{figure}[t]
\centering

\includegraphics[width=\linewidth]{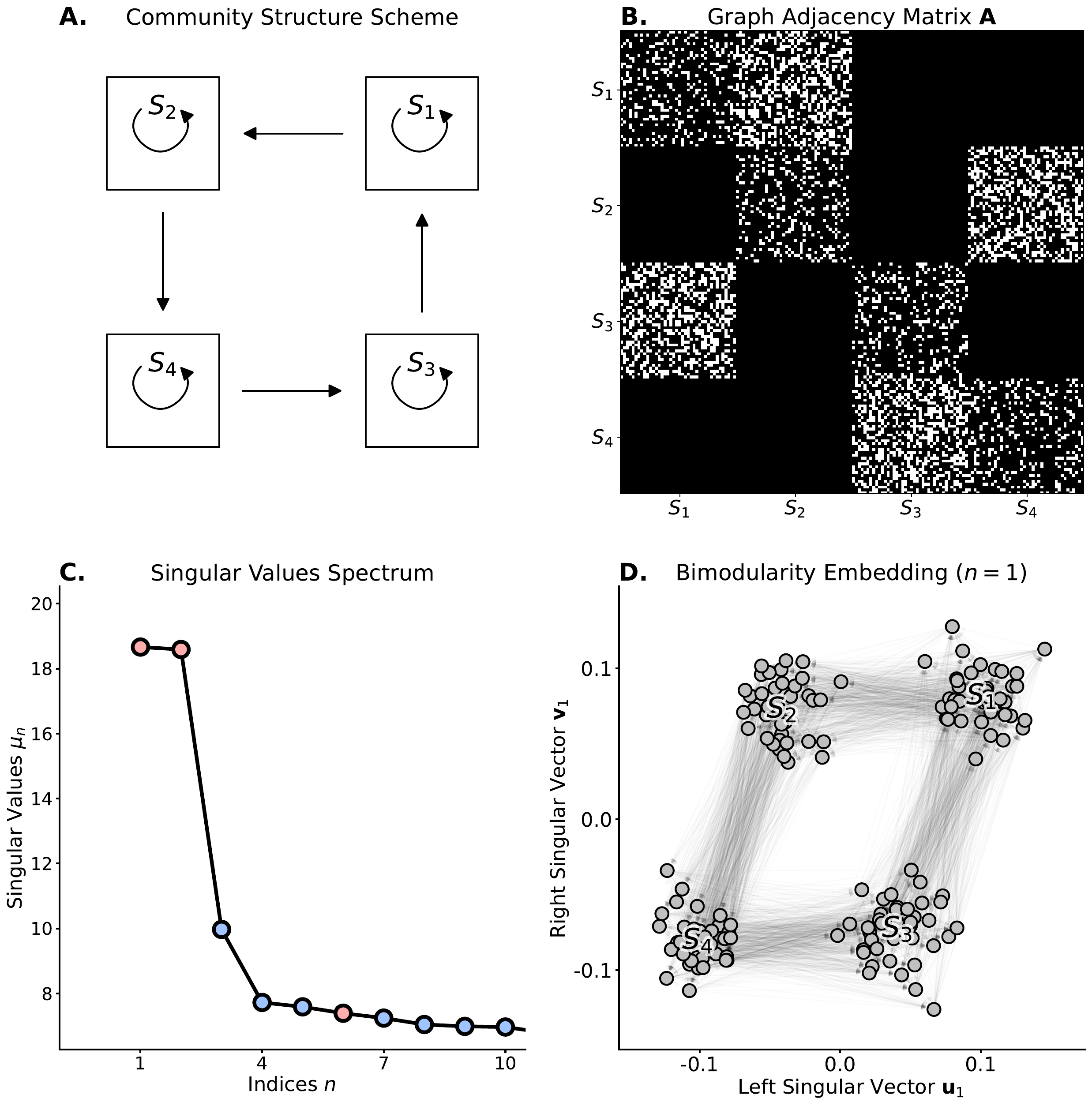}
\caption{
    Bimodular community structure.
    (A) Model of conventional communities ($S_1, S_2, S_3, S_4$) that additionally project onto one other according to a cycle structure.
    (B) Adjacency matrix $\mathbf{A}$ of a stochastic block graph of 200 nodes (50 per community) with the structure of A.
    (C) Spectrum of the 10 highest singular values of the singular value decomposition (SVD) of $\mathbf{B}$.
    Colors indicate the sign of the singular value associated to assortative (red, $\mu>0$) or dissortative (blue, $\mu < 0$) community structures.
    (D) Projection of the graph nodes on the two dimensional space given by the first component of the SVD of $\mathbf{B}$; i.e., the left and right singular vectors associated to the highest singular value $\mu_1$.
}
\label{fig:canonic}
\end{figure}

Let us consider the canonical directed community structure in Fig.~\ref{fig:canonic}.A; i.e., we have a  cycle between four sets of nodes that are not only densely connected within themselves, but also between them according to the directional pattern that is indicated.
Two types of communities can be recognized: first, the conventional community within the sets; second, the directed communities between the sets according to the cycle structure.
The adjacency matrix for an instantiation of this model is shown in Fig.~\ref{fig:canonic}.B.
Each set contains 50 nodes, connected with density $\gamma=30\%$ with randomly picked nodes within the set, and then connected to other sets with the same edge density and following the directions of the model.
The modularity matrix $\mathbf{B}$ is defined as $B_{ij} = A_{ij} - {k_i^\text{out} k_j^\text{in}}/{m}$.
The configuration model for this graph as well as the modularity matrix are illustrated in \SICanonicalModularity.

\subsection*{Spectral Method for Optimizing Bimodularity}
Finding the directed communities can be elegantly solved by a spectral method that maximizes the bimodularity index $Q_\text{bi}$.
If we revert for now to the bipartitioning problem where only two sending/receiving communities are to be identified, then this information can be encoded in two indicator vectors $\mathbf{s}^\text{out}$ and $\mathbf{s}^\text{in}$, respectively, of which the elements take values $+1/-1$ to indicate the partition.
The bimodularity index can then be rewritten as 
\begin{align}
    Q_\text{bi}(\mathbf{s}^\text{out}, \mathbf{s}^\text{in})&=
    \frac{1}{m} \sum_{\substack{(i,j) \text{ s.t.}\\\mathbf{s}^\text{out}[i]=\mathbf{s}^\text{in}[j]}} B_{ij} 
    = \frac{1}{2m} (\mathbf{s}^{\text{out}})^T\mathbf{B}\mathbf{s}^\text{in}.
\end{align}
Convex relaxation of the vectors $\mathbf{s}^\text{out}$ and $\mathbf{s}^\text{in}$ does not restrict their values to $+1/-1$ and instead imposes a unit norm, which allows to obtain the solution by the SVD: $\mathbf{B}=\mathbf{U\Sigma V}^{T}$; see Methods for the full derivation.
The columns of $\mathbf{U}$ contain the left singular vectors, those of $\mathbf{V}$ the right singular vectors, and the diagonal elements of $\mathbf{\Sigma}$ are the corresponding singular values that relate to bimodularity up to a factor $m$.

Akin to Laplacian spectral embedding, different pairs of singular vectors $(\mathbf{u}_k,\mathbf{v}_k)$ provide representations of the network by decreasing bimodularity \cite{von_luxburg_tutorial_2007, varshney_structural_2011}. For the running example, the singular values are plotted in Fig.~\ref{fig:canonic}.C and reveal two components with large bimodularity. The visualization of the first component in Fig.~\ref{fig:canonic}.D essentially unveils the diagram of the underlying model, with the four sets clearly separated and edges running counter-clockwise.

\begin{figure*}[ht]
\centering
\includegraphics[width=.9\linewidth]{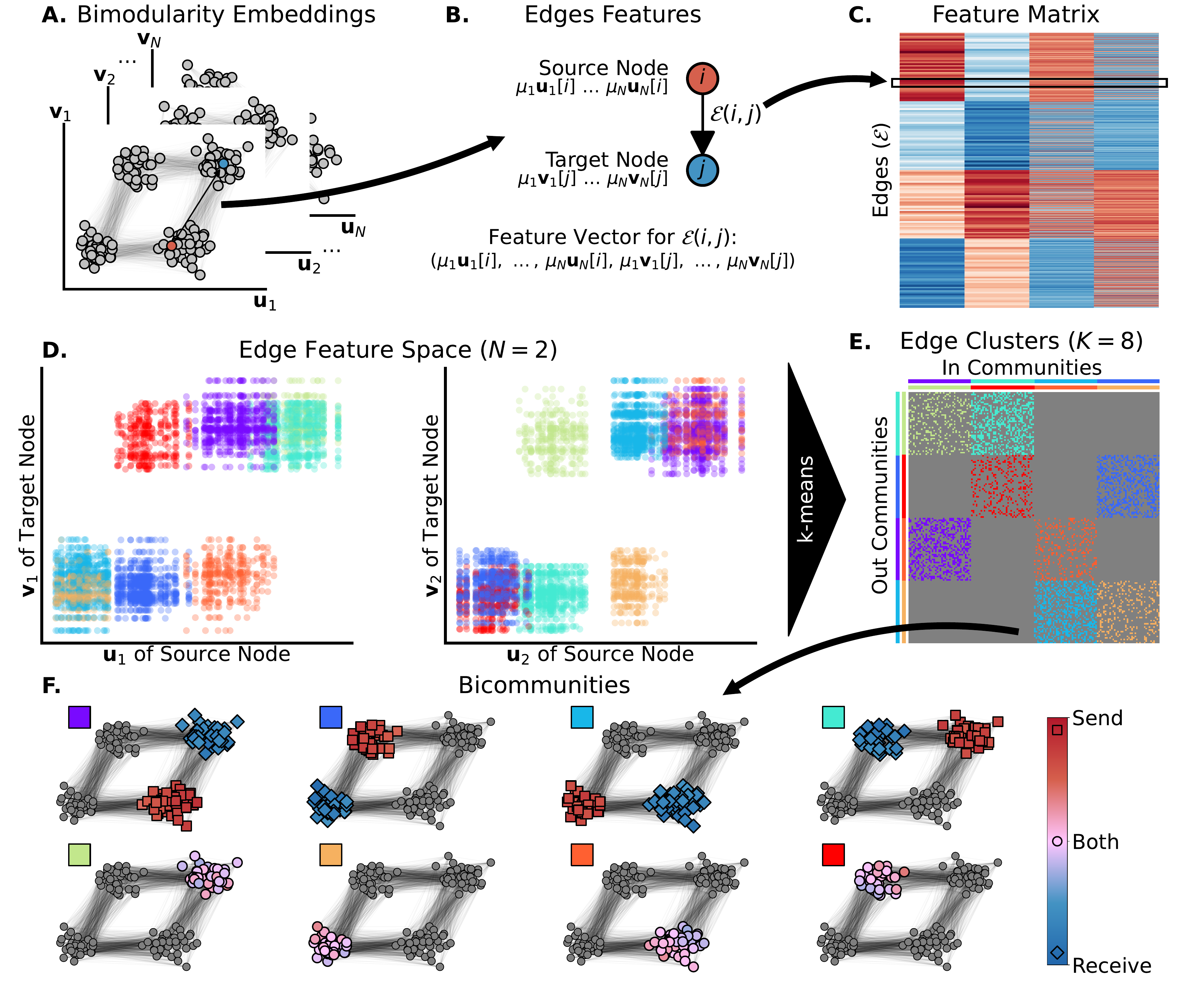}
\caption{
    Detection of bicommunities.
    (A) $N$ left $\mathbf{u}$ and right $\mathbf{v}$ singular vectors of the SVD of the modularity matrix $\mathbf{B}$ (bimodularity embeddings).
    (B) Aggregation of embeddings at the level of graph nodes to build a feature vector that represents the sending and receiving partition of the source ($i$) and target ($j$) node respectively.
    (C) Feature matrix with features of all graph edges.
    (D) Edge feature space (first two SVD components).
    Each dot represents an edge of the graph colored based on its corresponding cluster.
    (E) Graph adjacency matrix where each edge cluster is colored.
    Sending (left) and receiving (top) nodes are shown with colors of the corresponding cluster.
    (F) Bicommunities of the block-cycle graph with colors corresponding to the edge cluster.
    Colors and markers of nodes indicate whether a node belongs to the sending (red, square), receiving (blue, diamond) or both (circle, purple) community.
}
\label{fig:bicommunities}
\end{figure*}

\subsection*{Detection of Directed Communities}

Bimodularity-based embedding of the graph nodes provides insightful representations of network organization.
However, an additional step is needed to effectively detect pairs of sending $C_k^\text{out}$ and receiving communities $C_k^\text{in}$, termed \emph{bicommunities}.

The key to recover the mapping between sending and receiving communities from the embedding is to perform clustering of the edges instead of the nodes.
An edge is represented by the projection of its source node $i$ onto the sending partition and, likewise, the projection its target node $j$ onto the receiving partition.
For instance, the first component of the SVD provides for every edge $(i,j)$ the 2-D feature vector $(\mathbf{u}_1[i],\mathbf{v}_1[j])$.
For $N$ components, the complete feature vector for an edge $(i,j)$ can be obtained by concatenation and scaling by the singular values for stability purpose (see \SICanonicalStability) \cite{ng_spectral_2001}:
\begin{align*}
\mathbf{f}=\big(\mu_1 \mathbf{u}_1[i],\mu_1 \mathbf{v}_1[j],\ldots, \mu_N\mathbf{u}_N[i], \mu_N\mathbf{v}_N[j]\big).
\end{align*}
The set of all feature vectors is then fed into a clustering algorithm (i.e.,  k-means) to identify groups of edges that represent the bicommunity mappings.
The sending and receiving parts of such a bicommunity then emerge by considering the set of source and target nodes, respectively.
Note that, with this detection method, a sending community can overlap with its corresponding receiving partition (i.e., self-community) or with the sending or receiving part of other bicommunities.
The flowchart in Fig.~\ref{fig:bicommunities} illustrates the process for the running example.
The clustering of the edges leads to 8 communities, 4 conventional ones, and 4 bicommunities, thus retrieving the complete original model (Fig.\ref{fig:bicommunities}.E).

\subsection*{Bicommunities of the \textit{C. elegans} Neuronal Network}
We present bimodularity maximization to highlight directed communities in experimental data, in particular the neuronal connectome of the hermaphrodite nematode \textit{C. elegans} (see Methods).

\begin{figure*}[ht]
\centering
\includegraphics[width=\linewidth]{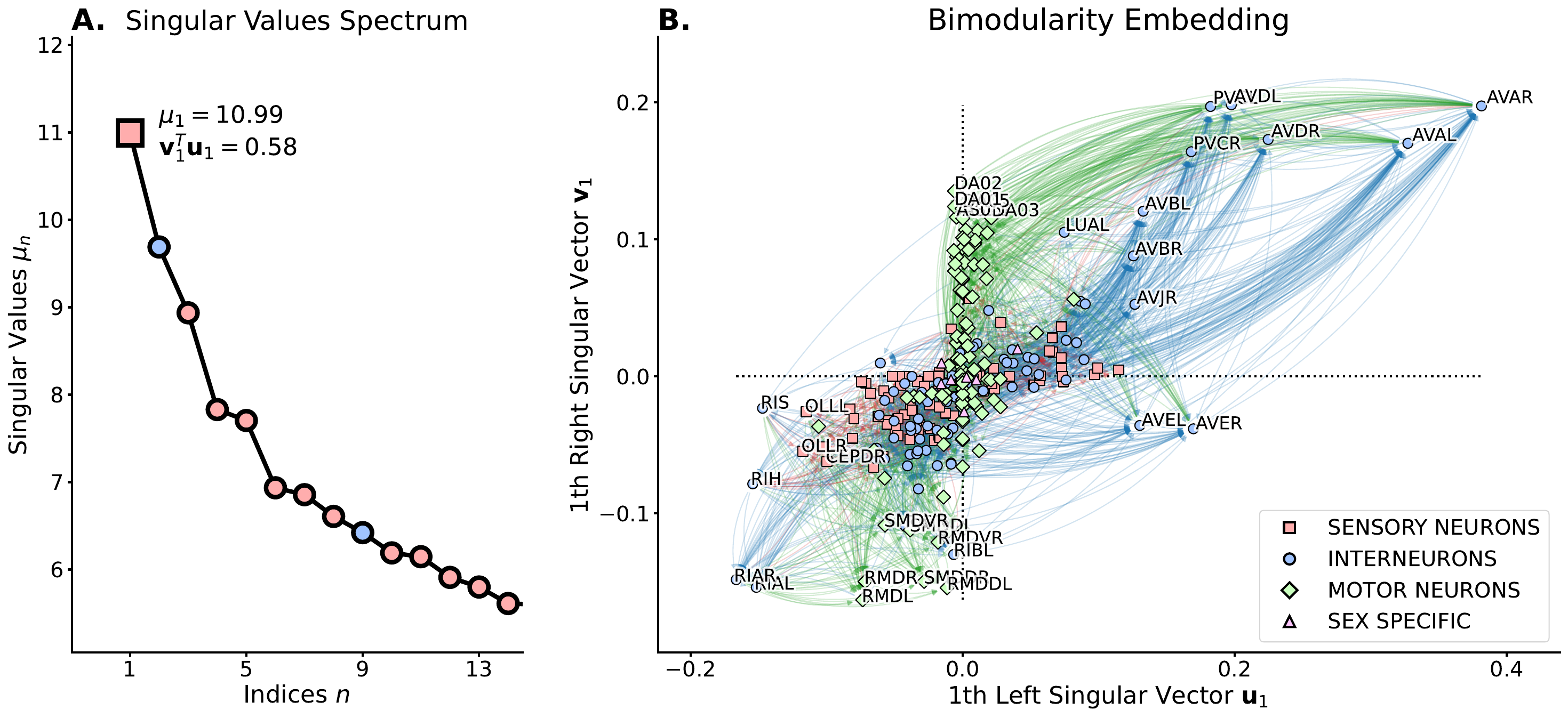}
\caption{
    Singular value decomposition (SVD) of the modularity matrix built from the \textit{C. elegans} wiring network of chemical synapses.
    (A) Spectrum of the singular values of the SVD.
    Colors indicate the sign of the singular value associated to assortative (red, $\mu>0$) or dissortative (blue, $\mu < 0$) community structures.
    (B) Projection of the graph nodes (neurons) onto the left ($\mathbf{u}_1$, horizontal axis) and right ($\mathbf{v}_1$, vertical axis) singular vectors of the bimodular component associated with the largest singular value ($\mu_1=10.99$).
    Colors and shape of nodes indicate the type of the neurons (sensory: red square; interneuron: blue circle; motor: green diamond; or sex specific: purple triangle).
    Edge colors indicate the type of the target node with the same color code as for the neuron types.
    Neuron labels are shown for graph nodes that are projected far from the origin $(0,0)$ of the axes.
}
\label{fig:worm}
\end{figure*}

Figure \ref{fig:worm}.A shows that there are several singular vectors with high bimodularity index.
The embedding of the neurons (nodes) onto the first bimodular component is displayed in Figure \ref{fig:worm}.B, which shows a distinct segregation between the head and body motion systems (see \SICelegansPos).
The head motion circuit appears in the lower-left quadrant ($\mathbf{u}_1<0$ and $\mathbf{v}_1<0$) of the embedding space with sensory (\textit{OLL, CEPD}) neurons projecting to intermediate ones (\textit{RIA, RIH, RIS}) and then reaching motor neurons (\textit{RMD, SMD}).
Similarly, neurons of the body motion circuit are embedded in the upper-right quadrant ($\mathbf{u}_1>0$ and $\mathbf{v}_1>0$).
Body motion neurons are distributed along the mid-upper part of the vertical axis ($\mathbf{u}_i=0$) and receive mostly from sensory (\textit{PHB, PLM}) and inter- (\textit{PVC, LUA}) neurons of the body.
Some notable head interneurons (\textit{AVA, AVB, AVD}) stand out within the body motion circuit and seem to provide major projections to both body inter- and motor neurons.
The \textit{AVE} interneurons seem to play a similar mediation role.
This complex interaction fully emerges from the analysis of directed connectivity through the lens of bimodularity. 

To further interpret these findings in terms of information pathways, we first denote the difference between sensory and motor neurons that tend to be more distributed on the horizontal and vertical axis of the embedding, respectively.
This is coherent with the idea that sensory (motor) neurons may have more edges going out (in) thus they will be aligned along the axis that characterizes the sending $\mathbf{u}_1$ (receiving $\mathbf{v}_1$) behavior.
With this in mind, we describe three information pathways (see \SICelegansPathways): the head motion pathway in the lower left quadrant; the body motion pathway in the upper right quadrant; and a last, more complex, pathway from head sensory to body motor neurons by passing through the key \textit{AVE} interneurons.
Specifically, these interneurons play a central part in conveying signals from the head system to the body system and finally back to motor neurons of the head and neck \SICelegansPathways.

\begin{figure}[t]
\centering
\includegraphics[width=\linewidth]{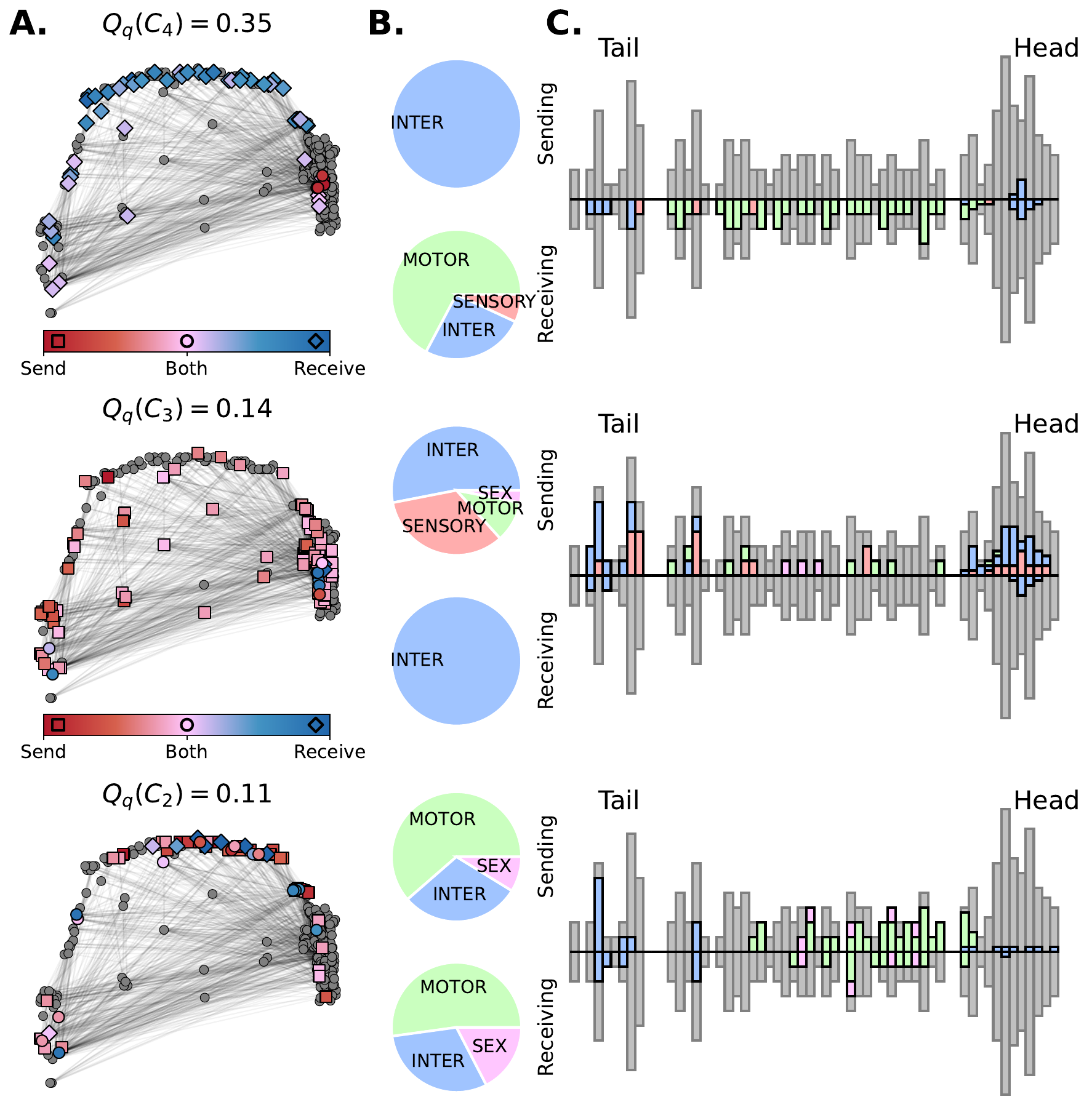}
\caption{
    Bicommunity of the \textit{C. elegans} extracted from the $N=5$ first bimodularity embeddings (components of the SVD).
    Only the clusters with top 3 highest bimodularity index (out of $K=5$) are displayed.
    (A) The sending and receiving parts of each bicommunity are highlighted on the spatial location of neurons.
    Markers indicate whether a neuron belongs to the sending (red square), the receiving (blue diamond) or both (pink circle) parts of a bicommunity.
    The colors represent the difference between the proportion of edges in the sending and in the receiving partitions with red and blue for nodes that tend to send and receive more respectively.
    (B) The distribution of neuron types (sensory, inter, motor and sex-specific neurons) in each sending or receiving community is summarized in the pie charts (middle).
    (C) The spatial distribution of bicommunities along the tail to head axis is detailed in the histograms (right) as a proportion of the total number of neurons (gray bars).
    The sending and receiving patterns are separated in the upper and lower bars respectively.
    Colors indicate the type of neuron following the color scheme of the pie charts.
    The body and tail histograms are made bigger (twice their original size/number of neurons) for visualization purpose.
}
\label{fig:worm-bicom}
\end{figure}

Figure \ref{fig:worm-bicom} shows the 3 bicommunities of the \textit{C. elegans} with highest bimodularity index (out of 5) extracted from the 5 first components of the SVD of $\mathbf{B}$.
These clusters highlight specific information pathways in the worm's motion system.
In detail, $C_3$ identifies communication from body sensory processes to key interneurons in the worm's brain (\textit{AVA}, \textit{AVB}, \textit{AVD}).
$C_4$ describes feed-forward transmissions of information from these interneurons to motor neurons of the body and interneurons of the tail (\textit{PVC}).
$C_2$ finally shows interconnectivity between motor neurons of the body and the neck as well as their communication with interneurons of the worm's tail.
Finer definitions of the \textit{C. elegans}' bicommunities is achieved with a larger number of edge clusters, see \SICelegansMoreBicom.

\section*{Discussion}

We investigated community structure in directed graphs and how the concept of sending and receiving communities is key to provide insight into such network organization.
Specifically, we revisited the definition of modularity for directed graphs to introduce bimodularity being tailored to account for the directed nature of edges.
The main idea behind bimodularity is to provide pairs of sending and receiving communities that are not necessarily mutually exclusive. 
We presented a spectral method for maximization of bimodularity through the SVD of the modularity matrix.
This optimization revealed embeddings based on bimodular structure.
We then proposed the appropriate clustering approach to combine these embeddings and obtain the bicommunities in terms of sending/receiving communities and their mappings.
We demonstrated the feasilibity of the approach on a synthetic models as well as to unravel the direct communication pathways in the C. elegans.
Neurons and connections which are central in the worm's sensory and motion systems are made visible in the first dominant bimodularity embedding.
Bicommunities dissect the sensory to motor feed-forward loop through key interneurons of the head and the tail.

Methods previously tackled the embedding of directed graph nodes through the SVD of graph operators (e.g. adjacency or Laplacian) \cite{sussman_consistent_2012,  rohe_co-clustering_2016, wang_spectral_2020, dang_community_2023}.
These approaches however usually fail to provide an interpretation of the embedding space.
This limitation is further emphasized as they do not leverage the intrinsic correspondence between the left and right singular vectors through the associated singular value.
The proposed bimodularity perspective is novel as it intuitively describes the relationship between a sending and its corresponding receiving projection.
Then, the bimodular edge embedding is a natural approach to identify edges mapping from nodes with similar sending patterns to nodes with similar receiving ones.
In comparison to conventional approaches, the proposed edge embedding is novel and unique to bimodularity in the interpretability of edge features \cite{evans_line_2009, ahn_link_2010}.

\added[id=AC]{
Advances in machine learning have recently also contributed to directed community detection \cite{chenSupervised2019, zhang_magnet_2021, khosla_node_2020}.
Specifically, graph neural networks that incorporate the adjacency line graph (i.e., edges of the original graph are turned into nodes and then connected when sharing an original node) \cite{chenSupervised2019}, or a spectral convolution layer via the magnetic Laplacian that is Hermitian \cite{zhang_magnet_2021, furutani_graph_2020}, have proven to be promising avenues for learning node embeddings.
However, learning-based methods rely on training processes in high-dimensional parameter spaces that limit their generalizability and interpretability.
In addition, it is not straightforward to understand how directionality is exploited to represent communities in terms of sending or receiving constituents.
To overcome this shortcoming, a shallower neural architecture has been proposed to learn node embeddings that represent the source and target roles \cite{khosla_node_2020}.
Specifically, single-layer neural models were used to extract embeddings from features of an alternating random walk.
Such identification of source and target roles is similar to our concept of sending and receiving communities, albeit there is no correspondence between sending and receiving patterns, as is the case for bimodularity.
These frameworks primarily focus on representation learning instead of unsupervised discovery.
In contrast, SVD-based edge embedding and detection of bicommunities flow naturally from the new definition of bimodularity.
Nevertheless, future work could integrate these operations into learning frameworks as well.
}

We investigated the properties of bimodularity, such as the fact that it coincides with conventional modularity when the graph is undirected (see Methods).
The separation of a graph into sending and receiving communities essentially augments the single dimension per component to two dimensions.
This allows to distinguish between nodes with high degree imbalance (sources and sinks) and to accurately place them in their respective sending/receiving partition (\SICanonicalSourceSink).
Specifically, we observe that only few source and sink nodes are required to create and identify sub-communities that receive or send more (for example followers and followed in a social network).
The combination of bimodularity embeddings into bicommunities is able to capture encoded structures and to accurately separate pre-encoded clusters in non-ideal examples that mimic real-world networks.

The application of the proposed scheme to the \textit{C. elegans}' neuronal wiring diagram further confirmed its relevance. In particular, the embedding according to the first component of the SVD encompasses the organization of the worm's head and body locomotion system and untangles information pathways such as head-to-head, body-to-body and head-to-body sensory-motor flows.
Specifically, this latter communication scheme is mainly driven by the \textit{AVE} neurons that could integrate information from the sensory/control neurons of the head system and further transmit signals to the body control and motor neurons.
We also observed that most body motor neurons also send information to more anterior motor neurons in the neck and the head, which confirms previous findings of intricate feedforward loop between the head and body motion systems \cite{morone_symmetry_2019, sohn_topological_2011, cook_whole-animal_2019}.
In short, the innovative aspect of bimodularity is that key clusters of neurons and their projections are captured in a single bimodular component (i.e., by a left/sending and right/receiving singular vectors) of the directed modularity matrix.

There is one methodological aspect that needs to be addressed when dealing with the SVD; i.e., the sign ambiguity singular values and vectors \cite{bro_resolving_2008}.
Indeed, in the $\mathbf{B=U\Sigma V}^T$ decomposition the sign of a singular value and one singular vector: $\mathbf{B=U\Sigma V}^T=\mathbf{U(-\Sigma) (-V}^T)$ can be freely swapped , which raises the need for a clear sign convention for univocal interpretation.
In the case of undirected graph, the modularity matrix is symmetric and the singular value (eigenvalue) hints whether a community structure is assortative (positive) or rather dissortative (negative, also known as anti-modular) \cite{snijders_estimation_1997, peixoto_bayesian_2019, zhang_statistical_2020}.
Here, we propose a sign convention associated to the dot product between the two singular vectors (i.e.,  cosine of the angle between them) and show that it is related to the assortative (or dissortative) nature of the partitions, see \SICanonicalDissort.
In particular, we observe that the bimodularity index of a bicommunity is negative when there are more edges between the sending and receiving sets of nodes than within them.
Therefore, this sign convention provides additional information about the observed community structure.
In the \textit{C. elegans} wiring network, the second component with the largest singular value is dissortative and supports the findings of the first component (Fig.~\ref{fig:worm}) by highlighting head interneurons that densely send information to body motor neurons without necessarily being connected (see \SICelegansSecEmbed\ for detailed description).

While bimodularity is a suitable metric for highlighting community structure in directed networks, as an extension of modularity it shares some of its limitations such as the resolution limit~\cite{fortunato_resolution_2007}.
Future work should investigate how this limit is influenced by directed information. 
In Supplementary Material (\SICanonicalBenchmark), we present a number of variations on the generated graphs that confirm the relevance and robustness of the bimodularity and bicommunity detection.
The presented method for bicommunity detection is parametrized by the number of bimodular components $N$ and the number of desired clusters $K$.
While $K$ is here tied to the clustering algorithm (i.e., k-means), we show that influence of $N$ is attenuated by multiplying entries in the edge feature (singular vectors $\mathbf{u}_k$ or $\mathbf{v}_k$) by their corresponding singular value $\mu_k$ (see \SICanonicalStability).
This is a common approach in spectral clustering to incorporate the spectrum of eigen-/singular values to weight the matrix of eigen-/singular vectors that is then fed to clustering \cite{von_luxburg_tutorial_2007, ng_spectral_2001}.

To conclude, we provide a theoretical framework based on bimodularity that extends and generalizes the community detection for directed graphs.
The joint optimization of the sending and the corresponding receiving communities unveils distinct pathways within networks that would remain invisible to conventional community detection.
Therefore, the new bimodularity index provides the foundation to better describe community structure in directed graphs and unlocks a whole new range of applications to networked data.

\matmethods{

All code and data used in this article are available as open-source code at \href{https://github.com/MIPLabCH/Bimodularity}{https://github.com/MIPLabCH/Bimodularity}.

\subsection*{Synthetic Graphs}

The synthetic graphs built and used in this article (such as the block cycle graph) have binary edge weights generated in a probabilistic manner following the stochastic block model (SBM) \cite{snijders_estimation_1997}.
We construct self-community blocks by creating edges with a probability $p_\text{self}$ and then assigning a direction to them with outgoing probability $p_\text{dir}=0.5$.
The resulting is a block-diagonal matrix of size $n_\text{blocks}\times n_\text{node per block}$.
Then these blocks are connected with the same approach, but by creating edges between self-communities with probability $p_\text{con}$ and assigning directions based on the desired structure (counter-clockwise for the block cycle graph).
Such models offered the possibility to freely adapt within- and between-communities edge densities ($p_\text{self}$ and $p_\text{con}$ respectively).
We emphasize that the model assigns a unique direction to an edge, thus preventing bi-directional relationships.
We however demonstrate that the presented approach is valid for networks composed of both uni- and bi-directional edges (as in the \textit{C. elegans}).
This constraint is made so that the symmetrized version of such a synthetic graph $\mathbf{A}_{undir}=\mathbf{A+A}^T$ will be a binary undirected graph with $p_\text{self}$ density of edges within communities and $p_\text{con}$ edge density between communities.
This means that if all self-communities are connected with density $p_\text{con}=p_\text{self}$ the symmetrized graph will be an unstructured random graph.

\subsection*{\textit{C. elegans} Wiring Network}

A directed binary graph was built from the wiring diagram of the \textit{C. elegans}' nervous system using information about 2194 chemical synapses between 279 neurons (graph nodes) \cite{varshney_structural_2011}.
Out of all these connections, 1961 (89\%) were asymmetric while only 233 (11\%) were bidirectional.
The undirected electric gap junction network was not used in the main analyses, but the bimodularity optimization and singular vector embedding of the joint (chemical synapses and gap junctions) graph are shown in \SICelegansGap.
Details about the type of each neuron (sensory, inter- or motor neurons) and their position in the worm's body have been gathered from \cite{cook_whole-animal_2019, brittin_volumetric_2018} and are summarized in \SIDatasetCelegans.

\subsection*{Bimodularity Index}
The definition of bimodularity is a measure of deviation of the edges configuration (and direction) compared to a null model.
Specifically, we leverage the idea that directed graphs may have different community structures when observing incoming or outgoing edges.
Hence, graphs with bimodular structure are expected to have more connections from sending communities to their corresponding receiving communities than expected ``on average''.
We makes use of the directed configuration model to express the null probability of having a directed edge between two nodes.
We then develop the expression of the bimodularity index:
\begin{align*}
    Q_\text{bi}&=\frac{1}{m} \sum_{k=1}^K \sum_{\substack{i\in C_{k}^\text{out}\\ j\in C_{k}^\text{in}}}\left[A_{ij}-\mathbb{E}(A_{ij}|\mathcal{H}_0)\right]\\
    &=\frac{1}{m}\sum_{k=1}^K \sum_{\substack{i\in C_{k}^\text{out}\\ j\in C_{k}^\text{in}}}\left[A_{ij}-\frac{k_i^\text{out} k_j^\text{in}}{m}\right]\\
    &=\frac{1}{m}\sum\limits_{i,j}\left[A_{ij}-\frac{k_i^\text{out} k_j^\text{in}}{m}\right]\delta_{C^\text{out},C^\text{in}}(i,j),
\end{align*}
where $\delta_{C^\text{out},C^\text{in}}$ is $1$ if $(i,j)$ runs between corresponding sending and receiving communities, and $0$ otherwise.

We consider the graph partition problem to separate the graph into two communities of nodes (sending community) that have common targets (receiving community).
Let $\mathbf{s}^\text{out}$ be a separator vector that takes value $s_i^\text{out}=1$ if node $i$ is in one sending partition and $s_i^\text{out}=-1$ if it is in the other.
Similarly, we define $\mathbf{s}^\text{in}$ as a separator vector for the receiving partition corresponding to the sending one.
Therefore, $\delta_{C^\text{out},C^\text{in}}(i,j)$ can be rewritten as
$$\frac{s_i^\text{out} s_j^\text{in}+1}{2}$$
that is $1$ if $s_i^\text{out}=s_j^\text{in}$ and $0$ otherwise.
We then develop the expression of bimodularity as:
\begin{align*}
    Q_\text{bi}&=\frac{1}{m}\sum\limits_{i,j}\left[A_{ij}-\frac{k_i^\text{out} k_j^\text{in}}{m}\right]\frac{s_{i}^\text{out} s_{j}^\text{in}+1}{2}\\
    &=\frac{1}{2m}\sum\limits_{i,j}\left(\left[A_{ij}-\frac{k_i^\text{out} k_j^\text{in}}{m}\right]s_{i}^\text{out} s_{j}^\text{in}+\left[A_{ij}-\frac{k_i^\text{out} k_j^\text{in}}{m}\right]\right).
\end{align*}
Knowing that $\sum\limits_{i,j}A_{ij}=m$, that $\sum\limits_i k_i^\text{out}=m$ and that $\sum\limits_j k_j^\text{in}=m$, the right part of the sum is equal to zero and the bimodularity index is can be expressed as a matrix multiplication:
\begin{align*}
    Q_\text{bi}&=\frac{1}{2m}\sum\limits_{i,j} \underbrace{\left[A_{ij}-\frac{k_i^\text{out} k_j^\text{in}}{m}\right]}_{=B_{ij}} s_{i}^\text{out} s_{j}^\text{in}\\
    &=\frac{1}{2m} (\mathbf{s}^\text{out})^T\mathbf{B} \mathbf{s}^\text{in}
    =Q_\text{bi}(\mathbf{s}^\text{out}, \mathbf{s}^\text{in}).
\end{align*}

For the undirected case, the definition of bimodularity falls back to the one of modularity. 
Indeed, the sending and receiving partitions $\mathbf{s}^\text{out}$ and $\mathbf{s}^\text{in}$ are the same when edges have no directions and the number of directed edges $m$ is twice the number of undirected edges $2n=m$:
\begin{align*}
    Q_\text{bi}(\mathbf{s}^\text{out}, \mathbf{s}^\text{in})&=\frac{1}{2m} (\mathbf{s}^\text{out})^T\mathbf{B} \mathbf{s}^\text{in}
    =\frac{1}{4n} \mathbf{s}^T \mathbf{B} \mathbf{s}=Q(\mathbf{s}).
\end{align*}

\subsection*{Optimization of Bimodularity} \label{methods:opti}
Maximization of bimodularity $Q_\text{bi}(\mathbf{s}^\text{out},\mathbf{s}^\text{in})$ is achieved by convex relaxation of the separator vectors $\mathbf{s}^\text{out}, \mathbf{s}^\text{in}$.
In essence, the separator vectors are not anymore restricted to $+1/-1$ but under the constraint to be unitary vectors such that $(\mathbf{s}^\text{out})^{T}\mathbf{s}^\text{out}=1$ and $(\mathbf{s}^\text{in})^{T}\mathbf{s}^\text{in}=1$.
We then write the unconstrainted optimization problem using Lagrange multipliers $\mu^\text{out}$ and $\mu^\text{in}$:
\begin{align*}
    \max_{\mathbf{s}^\text{out},\mathbf{s}^\text{in}} (\mathbf{s}^\text{out})^{T}\mathbf{B} \mathbf{s}^\text{in}+\mu^\text{out}(1-(\mathbf{s}^\text{out})^{T}\mathbf{s}^\text{out})+\mu^\text{in}(1-(\mathbf{s}^\text{in})^{T}\mathbf{s}^\text{in}).
\end{align*}
Partial derivatives with respect to $\mathbf{s}^\text{out}$ and $\mathbf{s}^\text{in}$ leads to:
\begin{align*}
    \left\{ \begin{array}{rcc}
    (\mathbf{s}^\text{in})^{T}\mathbf{B}^{T}-\mu^\text{out} (\mathbf{s}^\text{out})^{T}=0\\
    (\mathbf{s}^\text{out})^{T}\mathbf{B}-\mu^\text{in} (\mathbf{s}^\text{in})^{T}=0
    \end{array}\right. \rightarrow
    \left\{ \begin{array}{rcc}
    \mu^\text{out} \mathbf{s}^\text{out}=\mathbf{Bs}^\text{in}\\
    \mu^\text{in} \mathbf{s}^\text{in}=\mathbf{B}^{T}\mathbf{s}^\text{out}
    \end{array}\right.\,.
\end{align*}
Multiplying the first expression by $\mathbf{B}^T$,  we then obtain $\mu^\text{out}\mathbf{B}^T\mathbf{s}^\text{out} =\mathbf{B}^T\mathbf{Bs}^\text{in}$, where we can substitute the second expression to obtain
$$\mu^\text{out}\mu^\text{in}\mathbf{s}^\text{in} =\mathbf{B}^T\mathbf{Bs}^\text{in}.$$
Here we recognize that $\mathbf{s}^\text{in}$ is an eigenvector of the symmetric matrix $\mathbf{B}^T\mathbf{B}$ with the corresponding eigenvalue $\lambda^{\text{in}}=\mu^\text{out}\mu^\text{in}$.
Similarly, applying the same approach to the second (lower) expression and substituting the upper one, we obtain
$$\mu^\text{out}\mu^\text{in}\mathbf{s}^\text{out} =\mathbf{BB}^T\mathbf{s}^\text{out},$$
which shows that $\mathbf{s}^\text{out}$ is an eigenvector of $\mathbf{B}\mathbf{B}^T$ with the eigenvalue $\lambda^{\text{out}}=\mu^\text{out}\mu^\text{in}$.

We observe that the eigenvectors obtained from the eigendecomposition of $\mathbf{B}\mathbf{B}^T$ and $\mathbf{B}^T\mathbf{B}$ are equivalent to the left and right singular vectors of the singular value decomposition (SVD) of $\mathbf{B}$, respectively, with the eigenvalues $\lambda_i$ being the squared singular values $\mu_i^2$.
In particular, from the SVD 
$$ \mathbf{B}=\mathbf{U\Sigma V}^T,$$
we can derive
\begin{align*}
    \left\{ \begin{array}{rcc}
    \mathbf{B}\mathbf{B}^T=\mathbf{U\Sigma V}^T\mathbf{V\Sigma U}^T=\mathbf{U\Sigma}^2\mathbf{U}^T=\mathbf{U\Lambda}\mathbf{U}^T\\
    \mathbf{B}^T\mathbf{B}=\mathbf{V\Sigma U}^T\mathbf{U\Sigma V}^T=\mathbf{V\Sigma}^2\mathbf{V}^T=\mathbf{V\Lambda}\mathbf{V}^T
    \end{array}\right.\,,
\end{align*}
meaning that $\mathbf{s}^\text{out}$ and $\mathbf{s}^\text{in}$ can be identified as the columns of $\mathbf{U}$ and $\mathbf{V}$, respectively.
Therefore, when choosing $\mathbf{s}^\text{out}=\mathbf{u}_i$ and $\mathbf{s}^\text{in}=\mathbf{v}_i$ the bimodularity index is proportional to corresponding to the singular value (up to a normalization $\frac{1}{2m}$):
\begin{align*}
    Q_\text{bi}(\mathbf{u}_i, \mathbf{v}_i)=\frac{1}{2m}\mathbf{u}_i^T\mathbf{Bv}_i=\frac{1}{2m}\mathbf{u}_i^T\mathbf{U\Sigma V}^T\mathbf{v}_i=\frac{\mu_i}{2m}.
\end{align*}

\subsection*{Detection of Bicommunity} \label{methods:detect}
We present an elegant method to extract a set of sending and their corresponding receiving communities (i.e., bicommunities) by combining $N\geq1$ components of the SVD of $\mathbf{B}$.
For each graph edge, we aggregate the value of the $N$ leading left singular vector for the source node and the value of the $N$ leading right singular vectors for the target node.
The feature vector for edge $\mathcal{E}(i,j)$ from node $i$ to node $j$ thus is:
\begin{align*}
\mathbf{f}=\big(\mu_1 \mathbf{u}_1[i],\mu_1 \mathbf{v}_1[j],\ldots, \mu_N\mathbf{u}_N[i], \mu_N\mathbf{v}_N[j]\big).
\end{align*}
Clustering edges with such a feature vector will group edges that go from nodes in similar sending partition(s) to nodes in similar receiving partition(s).
This reminds the idea of a mapping $\mathcal{M}$ from a sending community $C_k^\text{out}$ to its corresponding receiving community $C_k^\text{in}=\mathcal{M}(C_k^\text{out})$.
Intuitively, the clustering approach separates edges into a sequence of mappings $\mathcal{M}(C^\text{out}_k)=C^\text{in}_k,\,k=1,\ldots,K$, from which we can derive the sending and receiving part of each bicommunity.
While it is clear that the clusters of edges are not overlapping, the nodes corresponding to the sending/receiving part of a bicommunity can be overlapping.
}

\showmatmethods{} 

\acknow{This work was supported in part by the Swiss National Science Foundation (SNSF), Sinergia project ``Precision mapping of electrical brain network dynamics with application to epilepsy'', grant number 209470.}

\showacknow{} 


\bibliography{Research_report}

\begin{thebibliography}{10}

\bibitem{varshney_structural_2011}
LR Varshney, BL Chen, E Paniagua, DH Hall, DB Chklovskii, Structural {Properties} of the {Caenorhabditis} elegans {Neuronal} {Network}.
\newblock {\em\protect\JournalTitle{PLOS Computational Biology}} \textbf{7}, e1001066 (2011) Publisher: Public Library of Science.

\bibitem{barabasi_network_2011}
AL Barabási, N Gulbahce, J Loscalzo, Network medicine: a network-based approach to human disease.
\newblock {\em\protect\JournalTitle{Nature Reviews Genetics}} \textbf{12}, 56--68 (2011) Publisher: Nature Publishing Group.

\bibitem{vertes_simple_2012}
PE Vértes, et~al., Simple models of human brain functional networks.
\newblock {\em\protect\JournalTitle{Proceedings of the National Academy of Sciences}} \textbf{109}, 5868--5873 (2012) Publisher: Proceedings of the National Academy of Sciences.

\bibitem{fornito_connectomics_2015}
A Fornito, A Zalesky, M Breakspear, The connectomics of brain disorders.
\newblock {\em\protect\JournalTitle{Nature Reviews Neuroscience}} \textbf{16}, 159--172 (2015) Publisher: Nature Publishing Group.

\bibitem{bassett_network_2017}
DS Bassett, O Sporns, Network neuroscience.
\newblock {\em\protect\JournalTitle{Nature Neuroscience}} \textbf{20}, 353--364 (2017).

\bibitem{li_graph_2022}
MM Li, K Huang, M Zitnik, Graph representation learning in biomedicine and healthcare.
\newblock {\em\protect\JournalTitle{Nature Biomedical Engineering}} \textbf{6}, 1353--1369 (2022) Publisher: Nature Publishing Group.

\bibitem{girvan_community_2002}
M Girvan, MEJ Newman, Community structure in social and biological networks.
\newblock {\em\protect\JournalTitle{Proceedings of the National Academy of Sciences}} \textbf{99}, 7821--7826 (2002) Publisher: Proceedings of the National Academy of Sciences.

\bibitem{fortunato_community_2016}
S Fortunato, D Hric, Community detection in networks: {A} user guide.
\newblock {\em\protect\JournalTitle{Physics Reports}} \textbf{659}, 1--44 (2016).

\bibitem{rohe_co-clustering_2016}
K Rohe, T Qin, B Yu, Co-clustering directed graphs to discover asymmetries and directional communities.
\newblock {\em\protect\JournalTitle{Proceedings of the National Academy of Sciences}} \textbf{113}, 12679--12684 (2016) Publisher: Proceedings of the National Academy of Sciences.

\bibitem{pathak_hierarchy_2024}
A Pathak, SN Menon, S Sinha, A hierarchy index for networks in the brain reveals a complex entangled organizational structure.
\newblock {\em\protect\JournalTitle{Proceedings of the National Academy of Sciences}} \textbf{121}, e2314291121 (2024) Publisher: Proceedings of the National Academy of Sciences.

\bibitem{bender_asymptotic_1978}
EA Bender, ER Canfield, The asymptotic number of labeled graphs with given degree sequences.
\newblock {\em\protect\JournalTitle{Journal of Combinatorial Theory, Series A}} \textbf{24}, 296--307 (1978).

\bibitem{newman_structure_2003}
MEJ Newman, The {Structure} and {Function} of {Complex} {Networks}.
\newblock {\em\protect\JournalTitle{SIAM Review}} \textbf{45}, 167--256 (2003) Publisher: Society for Industrial and Applied Mathematics.

\bibitem{holland_stochastic_1983}
PW Holland, KB Laskey, S Leinhardt, Stochastic blockmodels: {First} steps.
\newblock {\em\protect\JournalTitle{Social Networks}} \textbf{5}, 109--137 (1983).

\bibitem{snijders_estimation_1997}
TA Snijders, K Nowicki, Estimation and {Prediction} for {Stochastic} {Blockmodels} for {Graphs} with {Latent} {Block} {Structure}.
\newblock {\em\protect\JournalTitle{Journal of Classification}} \textbf{14}, 75--100 (1997).

\bibitem{peixoto_bayesian_2019}
TP Peixoto, Bayesian {Stochastic} {Blockmodeling} in {\em Advances in {Network} {Clustering} and {Blockmodeling}}.
\newblock (John Wiley \& Sons, Ltd), pp. 289--332 (2019) Section: 11 \_eprint: https://onlinelibrary.wiley.com/doi/pdf/10.1002/9781119483298.ch11.

\bibitem{blondel_fast_2008}
VD Blondel, JL Guillaume, R Lambiotte, E Lefebvre, Fast unfolding of communities in large networks.
\newblock {\em\protect\JournalTitle{Journal of Statistical Mechanics: Theory and Experiment}} \textbf{2008}, P10008 (2008).

\bibitem{lambiotte_communities_2009}
R Lambiotte, P Panzarasa, Communities, knowledge creation, and information diffusion.
\newblock {\em\protect\JournalTitle{Journal of Informetrics}} \textbf{3}, 180--190 (2009).

\bibitem{newman_modularity_2006}
MEJ Newman, Modularity and community structure in networks.
\newblock {\em\protect\JournalTitle{Proceedings of the National Academy of Sciences}} \textbf{103}, 8577--8582 (2006) Publisher: Proceedings of the National Academy of Sciences.

\bibitem{newman_finding_2006}
MEJ Newman, Finding community structure in networks using the eigenvectors of matrices.
\newblock {\em\protect\JournalTitle{Physical Review E}} \textbf{74}, 036104 (2006) Publisher: American Physical Society.

\bibitem{evans_line_2009}
TS Evans, R Lambiotte, Line graphs, link partitions, and overlapping communities.
\newblock {\em\protect\JournalTitle{Physical Review E}} \textbf{80}, 016105 (2009) Publisher: American Physical Society.

\bibitem{ahn_link_2010}
YY Ahn, JP Bagrow, S Lehmann, Link communities reveal multiscale complexity in networks.
\newblock {\em\protect\JournalTitle{Nature}} \textbf{466}, 761--764 (2010) Publisher: Nature Publishing Group.

\bibitem{arenas_size_2007}
A Arenas, J Duch, A Fernández, S Gómez, Size reduction of complex networks preserving modularity.
\newblock {\em\protect\JournalTitle{New Journal of Physics}} \textbf{9}, 176 (2007).

\bibitem{leicht_community_2008}
EA Leicht, MEJ Newman, Community {Structure} in {Directed} {Networks}.
\newblock {\em\protect\JournalTitle{Physical Review Letters}} \textbf{100}, 118703 (2008) Publisher: American Physical Society.

\bibitem{malliaros_clustering_2013}
FD Malliaros, M Vazirgiannis, Clustering and community detection in directed networks: {A} survey.
\newblock {\em\protect\JournalTitle{Physics Reports}} \textbf{533}, 95--142 (2013).

\bibitem{kim_finding_2010}
Y Kim, SW Son, H Jeong, Finding communities in directed networks.
\newblock {\em\protect\JournalTitle{Physical Review E}} \textbf{81}, 016103 (2010) Publisher: American Physical Society.

\bibitem{hosseini-pozveh_label_2022}
M Hosseini-Pozveh, M Ghorbanian, M Tabaiyan, A label propagation-based method for community detection in directed signed social networks.
\newblock {\em\protect\JournalTitle{Physica A: Statistical Mechanics and its Applications}} \textbf{604}, 127875 (2022).

\bibitem{dang_community_2023}
TD Dang, DH Do, THD Phan, Community detection in directed graphs using stationary distribution and hitting times methods.
\newblock {\em\protect\JournalTitle{Social Network Analysis and Mining}} \textbf{13}, 80 (2023).

\bibitem{wang_spectral_2020}
Z Wang, Y Liang, P Ji, Spectral {Algorithms} for {Community} {Detection} in {Directed} {Networks}.
\newblock {\em\protect\JournalTitle{Journal of Machine Learning Research}} \textbf{21}, 1--45 (2020).

\bibitem{sussman_consistent_2012}
DL Sussman, T ~, Minh, F ~, Donniell~E., , CE Priebe, A {Consistent} {Adjacency} {Spectral} {Embedding} for {Stochastic} {Blockmodel} {Graphs}.
\newblock {\em\protect\JournalTitle{Journal of the American Statistical Association}} \textbf{107}, 1119--1128 (2012) Publisher: ASA Website \_eprint: https://doi.org/10.1080/01621459.2012.699795.

\bibitem{von_luxburg_tutorial_2007}
U von Luxburg, A tutorial on spectral clustering.
\newblock {\em\protect\JournalTitle{Statistics and Computing}} \textbf{17}, 395--416 (2007).

\bibitem{ng_spectral_2001}
A Ng, M Jordan, Y Weiss, On {Spectral} {Clustering}: {Analysis} and an algorithm in {\em Advances in {Neural} {Information} {Processing} {Systems}}.
\newblock (MIT Press), Vol.{}~14, (2001).

\bibitem{chenSupervised2019}
Z Chen, L Li, J Bruna, Supervised community detection with line graph neural networks in {\em 7th international conference on learning representations, {ICLR} 2019}.
\newblock (2019).

\bibitem{zhang_magnet_2021}
X Zhang, Y He, N Brugnone, M Perlmutter, M Hirn, {MagNet}: {A} {Neural} {Network} for {Directed} {Graphs} in {\em Advances in {Neural} {Information} {Processing} {Systems}}.
\newblock (Curran Associates, Inc.), Vol.{}~34, pp. 27003--27015 (2021).

\bibitem{khosla_node_2020}
M Khosla, J Leonhardt, W Nejdl, A Anand, Node {Representation} {Learning} for {Directed} {Graphs} in {\em Machine {Learning} and {Knowledge} {Discovery} in {Databases}}, eds.{} U Brefeld, et~al.
\newblock (Springer International Publishing, Cham), pp. 395--411 (2020).

\bibitem{furutani_graph_2020}
S Furutani, T Shibahara, M Akiyama, K Hato, M Aida, Graph {Signal} {Processing} for {Directed} {Graphs} {Based} on the {Hermitian} {Laplacian} in {\em Machine {Learning} and {Knowledge} {Discovery} in {Databases}}, eds.{} U Brefeld, et~al.
\newblock (Springer International Publishing, Cham), pp. 447--463 (2020).

\bibitem{morone_symmetry_2019}
F Morone, HA Makse, Symmetry group factorization reveals the structure-function relation in the neural connectome of {Caenorhabditis} elegans.
\newblock {\em\protect\JournalTitle{Nature Communications}} \textbf{10}, 4961 (2019) Publisher: Nature Publishing Group.

\bibitem{sohn_topological_2011}
Y Sohn, MK Choi, YY Ahn, J Lee, J Jeong, Topological {Cluster} {Analysis} {Reveals} the {Systemic} {Organization} of the {Caenorhabditis} elegans {Connectome}.
\newblock {\em\protect\JournalTitle{PLOS Computational Biology}} \textbf{7}, e1001139 (2011) Publisher: Public Library of Science.

\bibitem{cook_whole-animal_2019}
SJ Cook, et~al., Whole-animal connectomes of both {Caenorhabditis} elegans sexes.
\newblock {\em\protect\JournalTitle{Nature}} \textbf{571}, 63--71 (2019) Publisher: Nature Publishing Group.

\bibitem{bro_resolving_2008}
R Bro, E Acar, TG Kolda, Resolving the sign ambiguity in the singular value decomposition.
\newblock {\em\protect\JournalTitle{Journal of Chemometrics}} \textbf{22}, 135--140 (2008) \_eprint: https://onlinelibrary.wiley.com/doi/pdf/10.1002/cem.1122.

\bibitem{zhang_statistical_2020}
L Zhang, TP Peixoto, Statistical inference of assortative community structures.
\newblock {\em\protect\JournalTitle{Physical Review Research}} \textbf{2}, 043271 (2020) Publisher: American Physical Society.

\bibitem{fortunato_resolution_2007}
S Fortunato, M Barthélemy, Resolution limit in community detection.
\newblock {\em\protect\JournalTitle{Proceedings of the National Academy of Sciences}} \textbf{104}, 36--41 (2007) Publisher: Proceedings of the National Academy of Sciences.

\bibitem{brittin_volumetric_2018}
CA Brittin, SJ Cook, DH Hall, SW Emmons, N Cohen, Volumetric reconstruction of main {Caenorhabditis} elegans neuropil at two different time points (2018) Pages: 485771 Section: New Results.

\end{thebibliography}

\includepdf[pages=-]{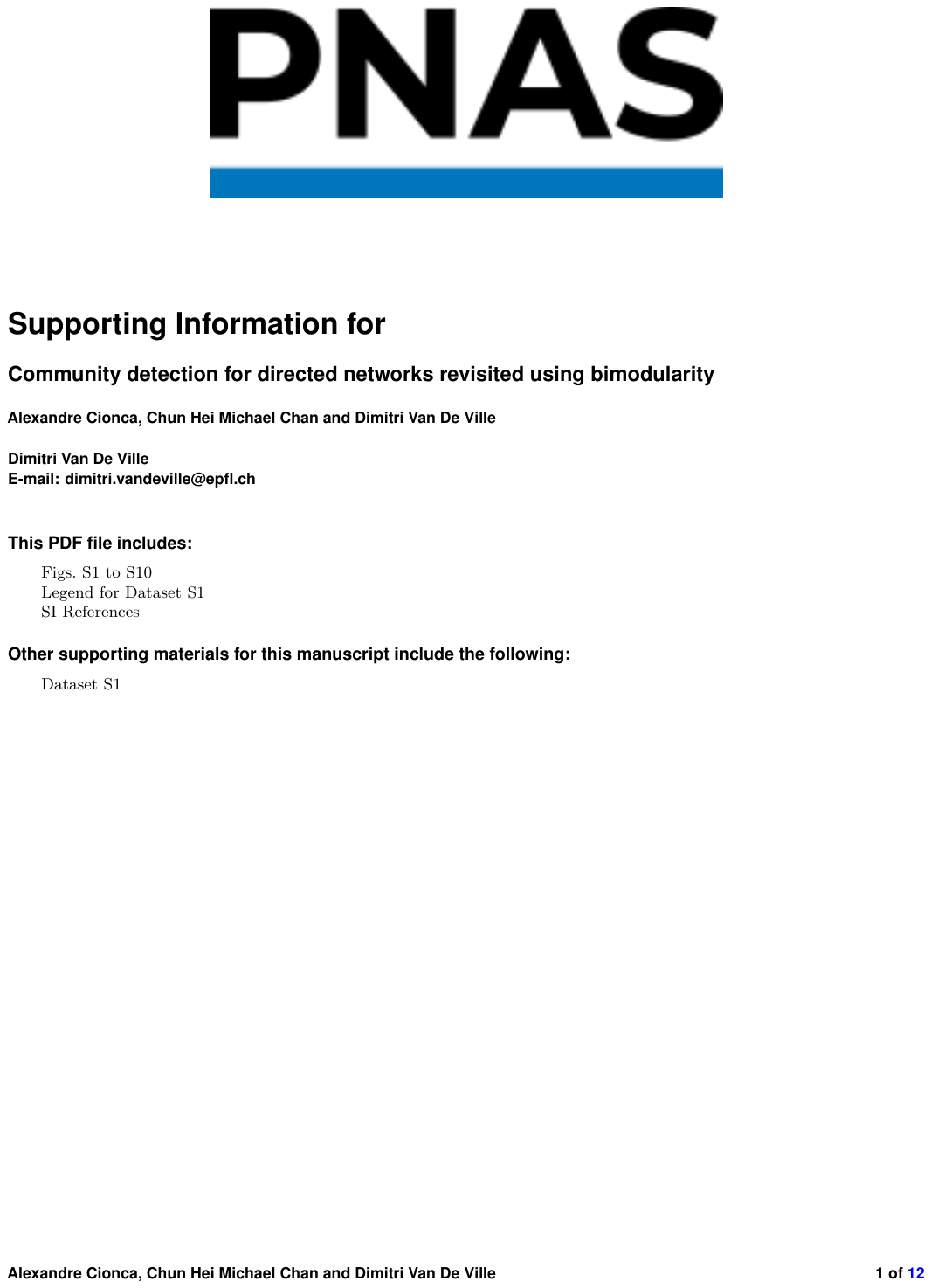}

\end{document}